\documentclass[a4paper]{article}
\usepackage{epsfig,amsfonts,amssymb,setspace}
\usepackage[T1]{fontenc}
\usepackage{amsmath}
\renewenvironment{thebibliography}[1]{\begin{oldthebibliography}{#1}\setlength{\parskip}{0ex}\setlength{\itemsep}{0ex}}{\end{oldthebibliography}}

\begin{document}
\fontsize{11}{11} \selectfont
\title{\bf Observations of Sy2 galaxy NGC\,3281\\ by XMM-Newton and INTEGRAL satellites}
\author{\textsl{A.\,A.\,Vasylenko$^{1}$\thanks{kvazarren@ukr.net}, E.\,Fedorova$^{2}$, V.\,I.\,Zhdanov$^{2}$}}
\date{\vspace*{-5ex}}
\maketitle
\begin{center} 
{\small  $^{1}$Faculty of Physics, Taras Shevchenko National University of Kyiv, Glushkova ave. 4, 03127 Kyiv, Ukraine\\
$^{2}$Astronomical Observatory, Taras Shevchenko National University of Kyiv, Observatorna str. 3, 04053 Kyiv, Ukraine\\}
\end{center}
\vspace*{-4ex}
\begin{abstract}
We present here the results of our analysis of X-ray properties of Seyfert 2 galaxy NGC\,3281, based on the observational data obtained by XMM-Newton and INTEGRAL within the energy ranges 0.2--12\,keV and 20--150\,keV, respectively. The XMM-Newton spectrum of this object is presented for the first time. We show that fitting the X-ray spectrum of this galaxy with models based on the reflection from the disc with infinite column density yields non-physical results. More appropriate fit takes into account both transmitted and reflected emission, passed through a gas-dusty torus-like structure. Keeping this in mind, to model the inhomogeneous clumpy torus, we used the {\tt MYTorus} model. Hence, we propose that the torus of NGC\,3281 is not continuous structure, but it consists of separate clouds, which is in a good agreement with the results of near-IR observations. Using this assumption, we found that the torus inclination angle and the hydrogen column density are $66.98^{+2.63}_{-1.34}$ degrees and $2.08^{+0.35}_{-0.18}\times10^{24}$\,cm$^{-2}$, respectively. Also, the emission of the hot diffuse gas with temperature $\sim590$\,eV and warm absorption were detected.\\[1ex]
\end{abstract}

\section*{\sc introduction}
\indent \indent Dusty toruses play a significant role in the Active Galactic Nuclei (AGN) Unification scheme \cite{AR} causing the main differences between various types of Seyfert galaxies. In obscured AGNs (Sy2) the nuclear emission is partially absorbed by a torus (depending on energy and column density, $N_{H}$, values). That is why the X-ray spectra of these objects are usually considered to contain two components: the emission reprocessed (scattered) by the dusty torus and the direct (unscattered) emission, which dominates at energies $\approx3-10$\,keV and $>10-12$\,keV, respectively. Thus the broad-band X-ray observations are often used to determine the AGN geometry and physical conditions of matter around a supermassive black hole (SMBH). Recent increasing of physical models based on Monte Carlo simulations provided us with a powerful tools for studying various properties of torus, such as geometry, column density, inclination and opening angles. The NGC\,3281 galaxy is very interesting object in this regard, because it is a candidate for Compton-thick source. It is is a near-by, low-redshifted ($z\approx0.010674$) X-ray bright radio quiet, Sab spiral galaxy with type 2 active galactic nucleus (Sy2). ASCA observations of this object revealed relatively hight value of hydrogen column density $N_{H}\approx 7-9\cdot10^{23}$\,cm$^{-2}$ \cite{Simpson}, taking into account that in the mid-infrared (mid-IR) amplitude $A_{V}\approx22^{m}$. Thus, following \cite{Simpson}, $N_{H}/A_{V}$ is a value of the order of magnitude significantly larger than the one determined along the line of sight in the Milky Way \cite{BSD}. That fact can be considered as a proof of the presence of the optically-thick absorber of both X-ray and IR emission here, possibly with partial covering. Vignali {\&} Comastri \cite{VC} studied NGC\,3281 galaxy in X-ray, treating the BeppoSAX data, and had shown that NGC\,3281 is a Compton-thick source with $N_{H}\approx1.5-2\cdot10^{24}$\,cm$^{-2}$. Their $N_{H}/A_{V}$ value is two times higher than that obtained by Simpson from ASCA observations. They also showed the presence of reflection dominated continuum at 3--10\,keV, with a relatively strong ($\text{EW}\approx0.5-1.2$\,keV) Fe\,K$\alpha $ emission line. Afterwards, NGC\,3281 is one of 153 galaxies of the Swift/BAT 9-month catalogue, and the Swift/BAT survey revealed here the hydrogen column density $N_{H}\approx8.6\cdot10^{23}$\,cm$^{-2}$ and the central black hole mass $\approx4\cdot10^{8}$\,M$_{\odot}$ \cite{WMRT}.

Sales et al. \cite{SPR} studied NGC\,3281 with mid-IR spectra obtained with the Gemini-South telescope. They found a very
deep silicate absorption and confirmed Compton-thick nature of NGC\,3281. For an explanation of spectroscopic properties they suggest that NGC\,3281 has a dusty clumpy torus with number of clouds in equatorial radius equal to 14 and inclination angle around $i=60^{\circ}$. Finally, De Rosa et al. \cite{RPB} obtained the value of $N_{H}\approx1.5\cdot10^{24}$\,cm$^{-2}$ using the BeppoSAX and INTEGRAL data.

In the present work, we use one of the new models of AGN torus, the {\tt MYTorus} model by Murphy and Yaqoob \cite{MY}, to describe the combined spectrum of the NGC\,3281 nucleus from INTEGRAL/ISGRI and recent XMM-Newton/EPIC data. Being compared with models which include the partial covering absorption with additional reflection component describing reflection for Compton-thick torus \cite{RPB,VC}, this model should give us more physically reasonable parameters of dusty torus.

\section*{\sc observations and data reduction}
\begin{figure}
\centering
\begin{minipage}[t]{.99\linewidth}
\centering \epsfig{file = 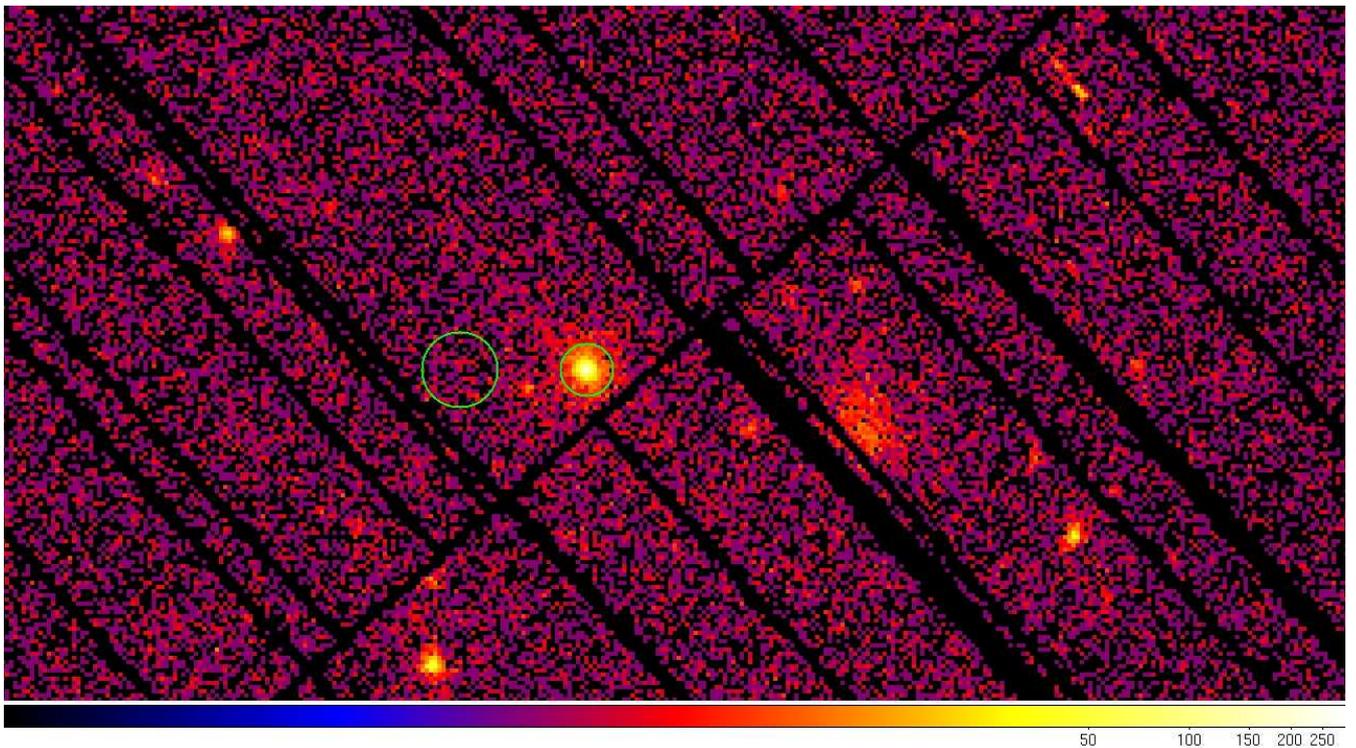,width = .99\linewidth}
\caption{The X-ray image of NGC\,3281, obtained by XMM-Newton EPIC PN camera (0.2--15\,keV).}\label{fig1}
\end{minipage}
\end{figure}

\vspace*{-1ex}
\indent\indent To study X-ray emission of NGC\,3281 galaxy we used the 1 May 2011 XMM-Newton observations (ID0650591001, PI name: L.\,Bassani) during the exposure time of 23715\,s. Data from EPIC cameras were processed with XMM SAS (v.11.0), applying the {\tt epproc} chain with standard settings. To obtain the XMM-Newton spectrum we collected the photons from the source and background areas on the same CCD chip (see Fig.\,\ref{fig1}). After filtering the EPIC data set we using the M.\,Guainazzi script\footnote{\tt http://xmm.esac.esa.int/sas/current/documentation/threads/epic\_merging.shtml} to combine the spectra from 3 EPIC cameras into one single spectrum in 0.2--12.0\,keV range.

After that, the effective duration of observations was 21848\,s with count-rate of 0.24\,cts/s. Spectra are rebinned to have at least 20 counts in each spectral channel. The analysis of the reflection grating spectrometer (RGS) spectrum is not reported here. We also analysed all 186 pointed public observations (on February 2013) by INTEGRAL/ISGRI for NGC\,3281 with effective exposure time 341\,ks with count-rate 0.9\,cts/s. ISGRI data analysis was performed using standard procedures of {\tt OSA~9.0}, the metatask {\tt ibis{\_}science{\_}analysis}. For fitting procedure we used the IBIS/ISGRI standard response matrix with 13 spectral channels. Thus, we obtained the good quality broad band spectrum in the range of 0.2--150\,keV. All spectral fits were performed using {\tt XSPEC}~v.12.6.0. The value of the intercalibration constant used in our simultaneous modelling of XMM-Newton/EPIC and INTEGRAL/ISGRI spectrum was in the range 1.2--0.9, i.\,e., within the typical range of values 2.0--0.5 \cite{RBU}.

\section*{\sc spectral analysis}
\vspace*{-1ex}
\indent\indent To reproduce the spectrum we applied two spectral models. The first model (model A hereafter) is based on  exponentially cut-off power-law spectrum reflected from neutral material ({\tt pexrav} \cite{MZ}). The second one (model B) is based on the toroidal model  of the Compton-thick X-ray reprocessor ({\tt MYTorus}\footnote{\tt http://mytorus.com/mytorus-manual-v0p0.pdf} \cite{MY}).  For our modelling we used the frozen value of the Galactic absorption, taken from Dickey \& Lockmann's work \cite{DL}, i.\,e. $6.42\cdot10^{20}$\,cm$^{-2}$. We assumed a cosmological framework with $H_{0}=70$\,km/s\,Mpc$^{-1}$, $\Lambda_{0}=0.73$, $\Omega_{M}=0.27$.

\subsection*{\sc spectral fits with model A}
\vspace*{-1ex}
\indent\indent The model A includes the following components: Galactic absorption (modelled by {\tt phabs} model in {\tt XSpec}), intrinsic partially covered absorption with power law ({\tt po$\times$zpcfabs}), thermal emission at soft X-rays and exponentially cut-off power-law spectrum reflected from neutral material of infinite cold slab ({\tt pexrav} \cite{MZ}), plus the Gaussian Fe\,K$_{\alpha}$ and Ni\,K$\alpha$ emission lines. The redshifted partial covering model (\texttt{zpcfabs}) is a multiplicative model defined as $M(E)=f\times\exp\left[-N_{H}\sigma\left(E[1+z]\right)\right]+(1-f)$. Here, $f$ is the covering fraction, $N_{H}$ is the neutral hydrogen column density, and $\sigma(E)$ is the photo-electric cross-section. A fully covering model (with $f=1$) corresponds to direct emission. The absorbed power-law model characterizes the nuclear component transmitted through a medium with  column density $N_{H}$. $A_{refl}$ is the normalization of the reflected component.

\begin{figure}
\centering
\begin{minipage}[t]{.98\linewidth}
\centering \epsfig{file = 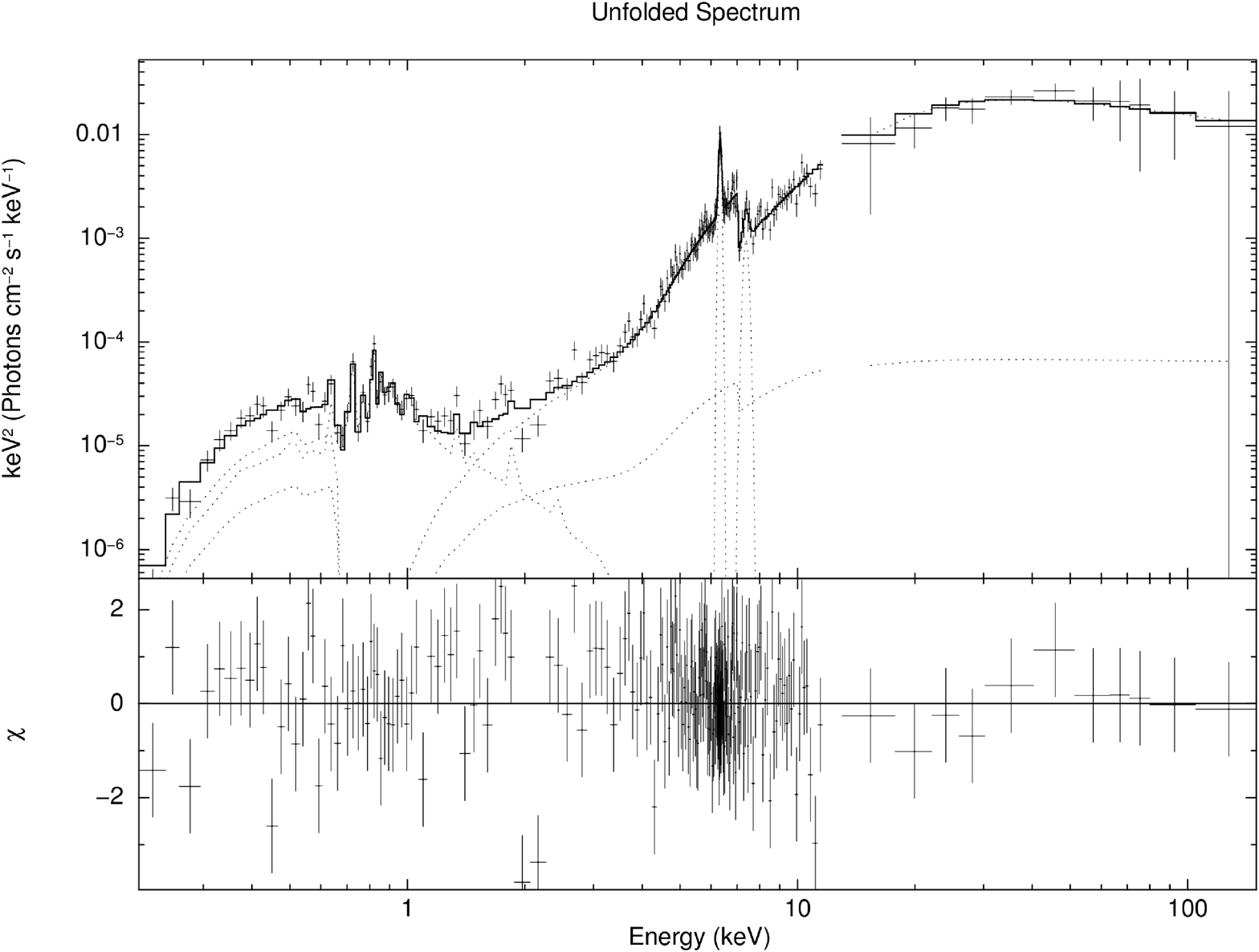,width = .9\linewidth}
\caption{NGC\,3281 unfolded broad-band spectrum with disk-reflection continuum.}\label{fig2}
\end{minipage}
\end{figure}

Model A demonstrates a quite good quality of fit with $\chi ^{2}$/$\nu=1.08$. The fitting parameters of the model A are shown in Table~\ref{tab1}, and in Fig.~\ref{fig2} the spectrum is shown. We have to point out that the obtained column density $N_{H}=5.38\cdot10^{23}$\,cm$^{-2}$ indicates that the galaxy is probably mildly obscured Sy2 \cite{Ric}, but not a Compton-thick Sy2. Our $N_{H}$ value is close to the value, obtained by Winter et al. \cite{WMRT}. It is worth to note that the value of relative reflection  $R\approx166$ is unphysical if related only to the geometry $R=\Omega/2\pi$. Here, $\Omega $ is a solid angle. So, in this interpretation, $R$ lies between 0 and 1. Another explanation of high value of $R$ is that part of the direct emission is blocked by partially covered material with the transmission efficiency being $\sim1/R $ \cite{UET}. Particularly, reflected material is represented as a dusty torus. Additionally, note that a torus should be distributed in a clumpy structure, because torus with smooth uniform distribution cannot survive close to the AGN \cite{KB}. Besides, we also found thermal emission from hot diffuse gas described by {\tt mekal} model with $kT=590$\,eV, and warm absorber described by {\tt wndabs} model.

\begin{table}
 \centering 
 \caption{Parameters of spectrum with {\tt pexrav} continuum model, for which $L^{intr}_{2-10	\text{keV}}=1.84\cdot10^{42}$\,ergs/s, 
$L^{intr}_{20-100\text{keV}}=1.27\cdot10^{43}$\,ergs/s, $\chi^2/\text{d.o.f}= 232/214$}\label{tab1}
 \vspace*{1ex}
\fontsize{9}{14}\selectfont
 \begin{tabular}{cccc}
 \hline Component& Parameter& Unit& Value \\
\hline
mekal& kT& keV& (5.90$^{+0.43}_{-0.42}$)$\cdot $10$^{-1}$ \\
 & abund& & (1.27$^{+0.13}_{-0.12}$)$\cdot $10$^{-1}$ \\
 & norm& & (9.62$\pm $0.62)$\cdot $10$^{ - 5}$ \\
\hline
wndabs& nH& 10$^{22}$ cm$^{-2}$& 1.34$^{+0.28}_{-0.19}$ \\
 & WindowE& keV& (6.58$^{+0.36}_{-0.46}$)$\cdot $10$^{-1}$ \\
 zpcfabs& nH& 10$^{22}$ cm$^{-2}$& 53.76$^{+1.75}_{-1.63}$ \\
 & CvrFract& & $(93.0^{+0.5}_{-0.4})\cdot 10^{-2}$\\
\hline
 zpowerlw& PhoIndex& & 2.040$^{+0.016}_{-0.014}$ \\
 & norm$^a$& & (8.86$\pm $0.26)$\cdot $10$^{ - 5}$ \\
\hline
 Arefl& & & 2.87$\pm $0.08 \\
\hline
 pexrav& foldE& keV& 1$\cdot 10^{4}$(fixed) \\
 & Rel{\_}refl& & 166.39$\pm $5.67 \\
\hline
 zgauss& Line E& keV& 6.398$^{+0.011}_{-0.012}$ \\
& Sigma& keV& (5.03$^{+1.25}_{-1.99}$ )$\cdot $10$^{-2}$ \\
 & norm& & (5.62$\pm $0.56)$\cdot $10$^{-5}$ \\
\hline
 zgauss& Line E& keV& 7.449$^{+0.041}_{-0.094}$ \\
 & Sigma& keV& (1.06$^{+0.58}_{-0.81}$)$\cdot $10$^{-1}$ \\
 & norm& & (1.62$\pm $0.55)$\cdot $10$^{-5}$ \\
\hline
\end{tabular}
  \\
$^a$ norm=[photons\,cm$^{-2}$\,s$^{-1}$\,keV$^{-1}].$
\end{table}

In the energy range 6--8\,keV we found two emission features with centroid energies 6.39\,keV and 7.45\,keV, which we interpreted as Fe\,K$_{\alpha }$ and Ni\,K$_{\alpha }$ emission lines, respectively. The FWHM value for these lines are $\sim 5500$\,km/s for Fe\,K$_{\alpha}$  and $\sim11000$\,km/s for Ni\,K$_{\alpha }$, and the equivalent widths EW are 526$\pm $4\,eV and $251\pm130$\,eV for the same lines, respectively, that leads us to the idea that Fe\,K$_{\alpha }$ line is originated quite far away from the central black hole, e.\,g, in the gaseous/dusty torus, and that the place of emission Ni\,K$_{\alpha}$ line is located closer to the centre, i.\,e., near the inner torus wall.

Despite of the good statistics of the model A, we keep in mind that physically it is not an adequate model, because {\tt pexrav} is not appropriate to describe the torus reflection in S2 type AGNs. This model supposes that the reflecting material is absolutely opaque (optically thick) and has disk geometry, but the torus can be clumpy and inhomogeneous with more or less transparent parts. Moreover, the torus geometry is also not known. Thus the {\tt pexrav} fit parameters, despite of good statistics, could not be considered as a physical characteristics of the torus in AGNs like the one considered here \cite{Yaq}.

\subsection*{\sc spectral fits with model B}
\vspace*{-1ex}
\indent\indent The second model (model B), instead of {\tt pexrav}, has another primary continuum component --  toroidal model  of the Compton-thick X-ray reprocessor ({\tt MYTorus} \cite{MY,Yaq}), which describes reflected emission from dusty torus with different geometry. Our case of {\tt MYTorus} model includes zero-order (intrinsic) continuum (source emission) and dual Compton-scattered continuum. The spectrum from the back-side reflection, which reaches the
observer without further interception by any absorbing matter, is approximated using a $0^{\circ}${\tt MYTS} component, and scattered emission, which is not from the back-side reflection, is approximated using a $90^{\circ}${\tt MYTS} component. To describe the intrinsic continuum we have chosen {\tt CompTT}, which is an analytic model describing Comptonization of soft photons in a hot plasma, developed by Titarchuk \cite{Tit}. Such continuum is characterized by the temperature and optical depth of the Comptonizing plasma ($kT$ and $\tau $, respectively). We also replaced partially covered absorption by simple absorption {\tt phabs}.

\begin{figure}
\centering
\begin{minipage}[t]{.98\linewidth}
\centering \epsfig{file = 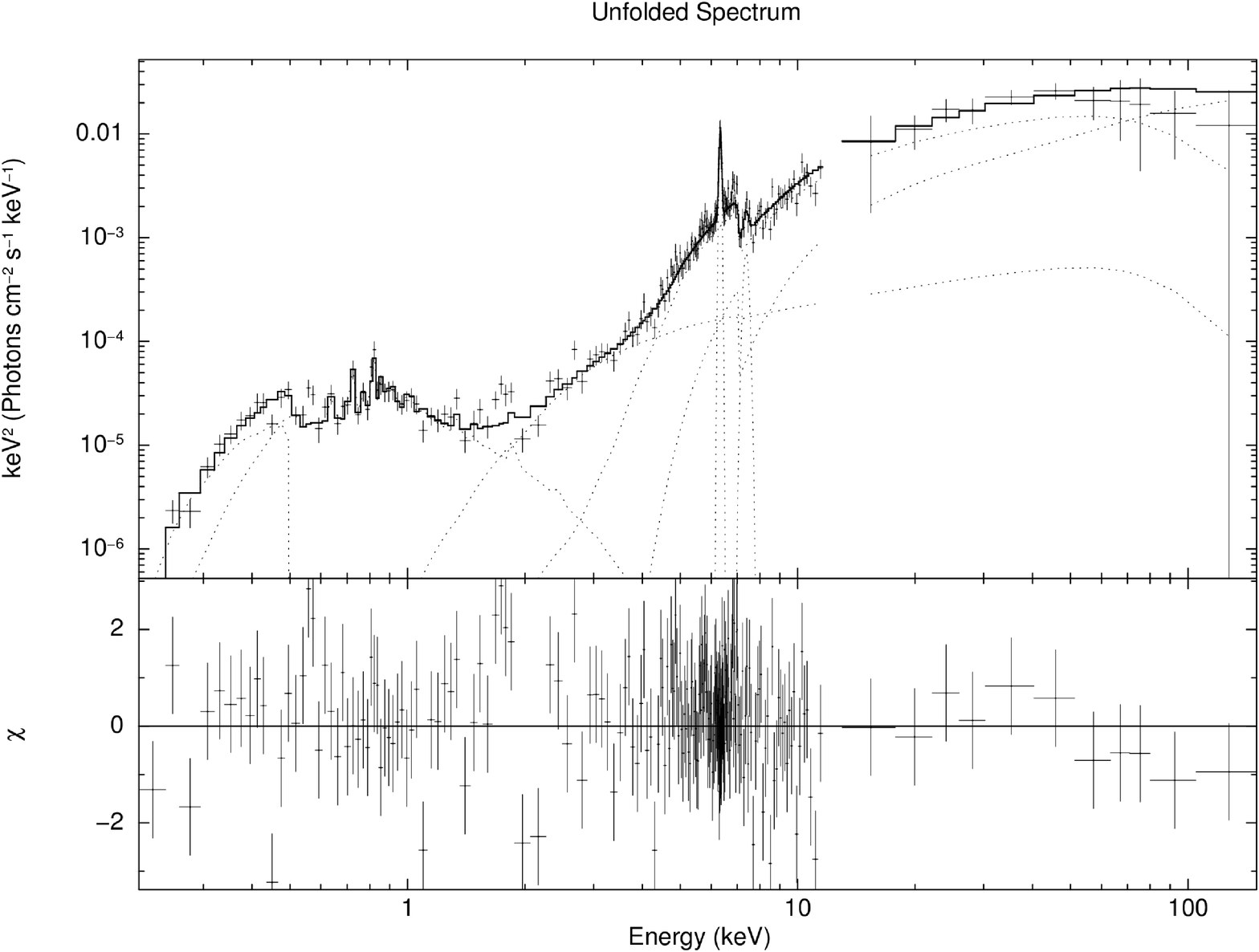,width = .9\linewidth}
\caption{NGC\,3281 unfolded broad-band spectrum with toroidal reprocessor model of continuum.}\label{fig3}
\end{minipage}
\end{figure}

The fitting parameters of the model B are presented in Table~\ref{tab2}, and in Fig.~\ref{fig3} the spectrum is shown.

\subsubsection*{\sc column densities}
\vspace*{-1ex}
\indent\indent We  obtained the value of the column density along the line of sight $N_{H}=2.08\cdot10^{24}$\,cm$^{-2}$. This value indicates that NGC\,3281 is a Compton-thick source, and it is comparable with the value, obtained by Vignali {\&} Comastri \cite{VC}. The obtained hydrogen column density for {\tt MYTorus} model are: $N_{H}=5.73\cdot10^{22}$\,cm$^{-2}$ for $0^{\circ}${\tt MYTS} and $N_{H}=5.99\cdot10^{23}$\,cm$^{-2}$ for $90^{\circ}${\tt MYTS}. Taking into account the meaning of these models we can conclude that the parameters of the Model~B properly define the spatial structure of the absorber. If we suppose that the X-ray source irradiates the torus isotropically then the hydrogen column density can be expressed in terms of Thompson optical depth \cite{Yaqoob}, which is $\tau_{T}=(11/9)N_{H}\sigma_{T}\approx0.81N_{24}$, thus in our case we have $\tau_{T}\approx2.08\cdot0.81\approx1.7$ along the line of sight.

\vspace*{-2ex}
\begin{table}
 \centering
 \caption{Parameters of spectrum with {\tt MYTorus} continuum model, for which $L^{intr}_{2-10\text{keV}}=5.56\cdot10^{41}$\,ergs/s, $L^{intr}_{20-100\text{keV}}=1.44\cdot10^{43}$\,ergs/s, $\chi^2/\text{d.o.f}=246/214$.}\label{tab2}
 \vspace*{1ex}
\fontsize{9}{12}\selectfont
 \begin{tabular}{cccc}
\hline Component& Parameter& Unit& Value \\
\hline mekal& kT& keV& (5.8$\pm $0.3)$\cdot $10$^{-1}$ \\
& abund& & (4.0$\pm $0.6) $\cdot $10$^{-2}$ \\
& norm& & (2.00$\pm $0.12)$\cdot $10$^{-4}$ \\
\hline wndabs& nH& 10$^{22}$ cm$^{-2}$& (5.4$^{+0.4}_{-0.3}$)$\cdot $10$^{-1}$ \\
& WindowE& kev& (6.58$\pm $1.12)$\cdot $10$^{-1}$ \\
\hline compTT& Tau& & 2.56$^{+0.56}_{-0.63}$ \\
& kT& keV& 40(fixed) \\
& norm$^a$& & (9.63$^{+0.28}_{-0.31}$)$\cdot $10$^{-5}$ \\
\hline MYTorusZ& nH& 10$^{24}$ cm$^{-2}$& 2.08$^{+0.35}_{-0.18}$ \\
& Incl& deg& 66.98$^{+2.63}_{-1.34}$ \\
\hline Const& & & 33.17$\pm $2.73 \\
\hline MYTorusS& nH& 10$^{24}$ cm$^{-2}$& (5.73$^{+0.48}_{-0.49}$) $\cdot $10$^{-2}$ \\
& Incl& deg& 0.0(fixed) \\
\hline Const& & & 92.61 $\pm $3.97 \\
\hline MYTorusS& nH& 10$^{24}$ cm$^{-2}$& (5.99$^{+0.30}_{-0.27}$)$\cdot $10$^{-1}$ \\
& Incl& deg& 90(fixed) \\
\hline zgauss& Line E& keV& 6.399$^{+0.012}_{-0.011}$ \\
& sigma& keV& (4.24$^{+1.2}_{-1.9}$)$\cdot $10$^{-2}$ \\
& norm& & (2.42$\pm $0.24)$\cdot $10$^{-5}$ \\
\hline zgauss& Line E& keV& 7.449$^{+0.104}_{-0.132}$ \\
& sigma& keV& (1.06$^{+1.46}_{-0.91}$)$\cdot $10$^{-1}$ \\
& norm& & (3.6$\pm $1.62)$\cdot $10$^{-6}$ \\
\hline
\end{tabular}
\\ $^a$ norm=[photons\,cm$^{-2}$\,s$^{-1}$\,keV$^{-1}].$
\end{table}

\subsubsection*{\sc intrinsic and scattered continua}
\begin{figure}
\centering
\begin{minipage}[t]{.98\linewidth}
\centering \epsfig{file = 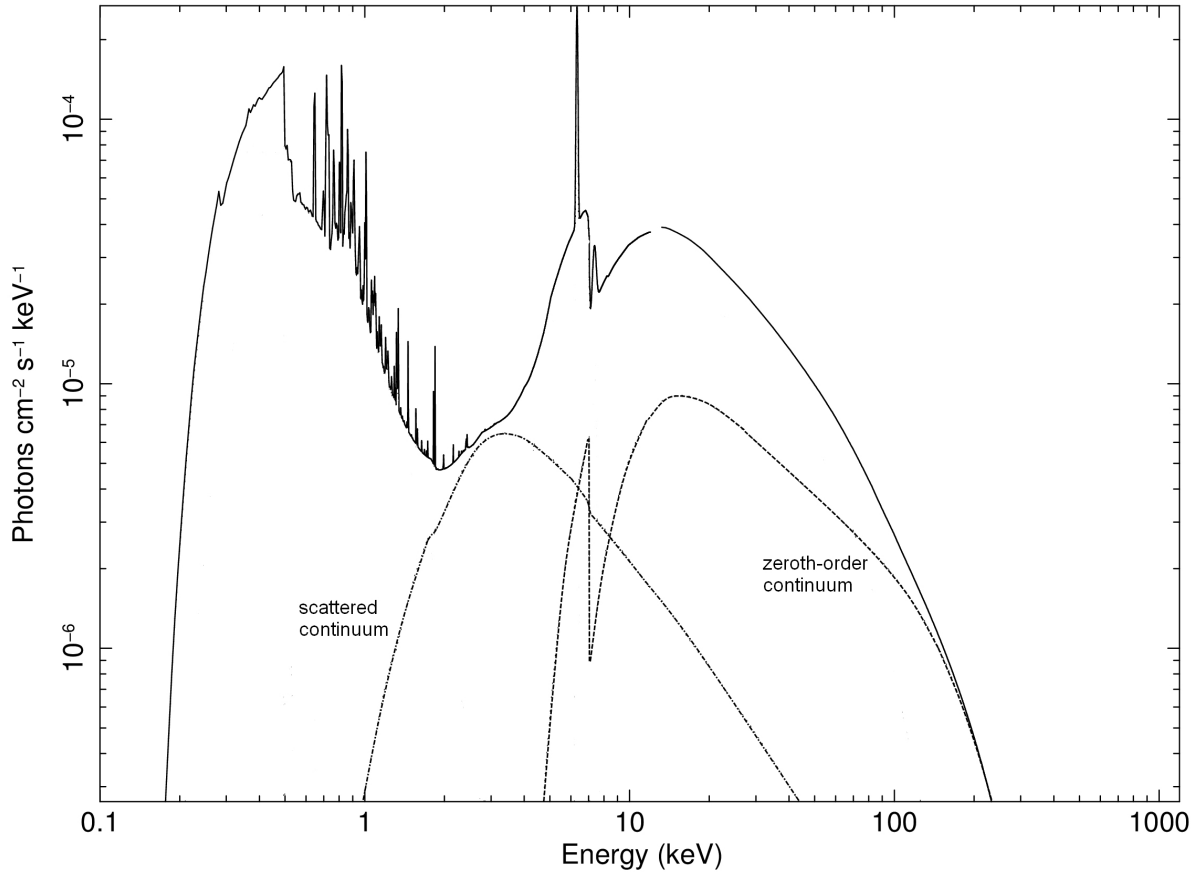,width = .9\linewidth}
\caption{The main components of NGC\,3281 model of broad-band spectrum with the toroidal X-ray reprocessor model of continuum. In addition to the total spectrum (solid line)  the zero-order continuum (dashed line) and the Compton-thick scattered spectrum from back-side reflection ($0^{\circ}${\tt MYTS} component, dash-dotted line) are shown.}\label{fig4}
\end{minipage}
\end{figure}

\vspace*{-1ex}
\indent\indent Using the {\tt compTT} model for the intrinsic continua, we can determine the plasma temperature and the optical depth, i.\,e. in our case $kT=40 $\,keV and $\tau \approx2.5$. Unfortunately, we cannot apply the models of reflected spectrum from accretion disc as intrinsic continua due to parameters degeneration for the model B.  But we have to note that the model B describes the spectrum above $\sim$70\,keV more worse than the model A. This can be due to the dominance of the disk reflection over the torus reflection on these energies. In support of this assumption, zeroth order continuum seen in Fig.\,\ref{fig4} dominates in the spectrum over the back-side reflection above $\sim10$\,keV. Taking also into account that the spectrum from the back-side reflection dominates over the inner torus wall emission below $\sim4$\,keV, we can conclude that we can observe the appearance of a clumpy dusty torus, but not a ``classical'' homogeneous one.

\subsubsection*{\sc Fe and Ni lines}
\indent\indent The {\tt MyTorus} model is consistent with the line emission in the Compton-thick medium. Thus within the Model~B the line parameters were being varied independently, despite the continuum model. So, we obtained the EW and FWHM values for  Fe\,K$_{\alpha }$ line EW\,$=589\pm54$\,eV. This is slightly higher than that obtained within the Model~A, but it is in a good agreement with the result of \cite{MY}. The FWHM value is $ \approx4600$\,km/s, which is also higher than the one obtained within the Model A. In the same time for the Ni\,K$_{\alpha }$ line, the normalization constant in this model is almost ten times higher and equivalent width is EW\,$= 176\pm73$\,eV. This value is in a good agreement with the models of Ni\,K$_{\alpha }$ line emission for our value of the hydrogen column density along the line of sight, described in \cite{YM}. We would like to stress that our discrepancy with the value of $\sim$2.3\,keV obtained by Vignali {\&} Comastri \cite{VC} can be induced by the fact that they neglected the influence of absorption on Fe\,K$\alpha$ line profile and intensity (see e.\,g. \cite{UET}.

\begin{figure}
\centering
\begin{minipage}[t]{.98\linewidth}
\centering \epsfig{file = 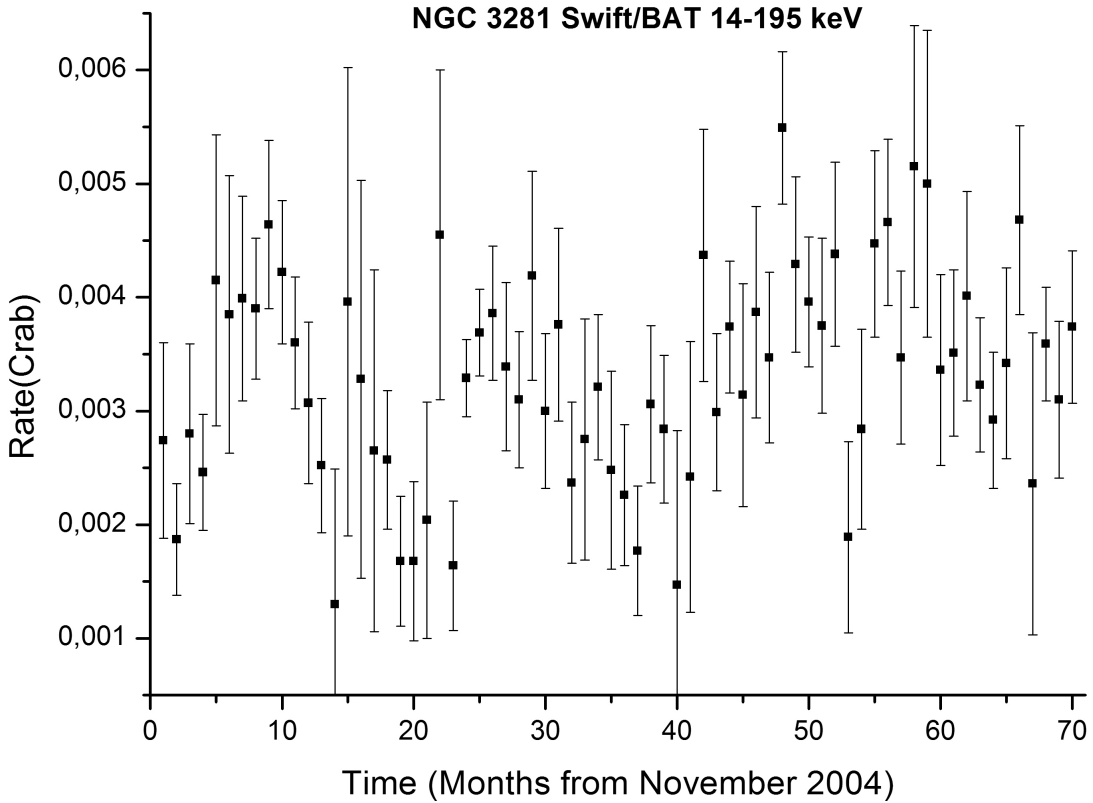,width = .85\linewidth}
\caption{NGC\,3281 Swift/BAT 70-month light curve in the 14--195\,keV. Each time bin corresponds to one month time interval, the first bin corresponding to 2004, November. The rate is given in ``Crab'' units.}\label{fig5}
\end{minipage}
\end{figure}

\section*{\sc discussion}
\vspace*{-1ex}
\indent\indent As we have mentioned in the introduction, in 1998 Simpson \cite{Simpson} found from the ASCA observations the ratio $N_{H}/A_{V}\approx3.2\cdot10^{22 }$\,cm$^{-2}$\,mag$^{-1}$ for NGC\,3281, which was large comparably to the Galaxy \cite{BSD}. Having considered the possible explanations of the near-IR emission, he rejected the compact star formation areas and optically thin clouds in the wide emission lines area. The best hypothesis he proposed was the optically thin medium containing thick clouds. Thus the torus surrounding the nucleus can be essentially inhomogeneous. Afterwards Sales et. al. \cite{SPR} confirmed the conclusions of Simpson, and additionally determined the inclination of the torus $i=60^{\circ}$. Note that this value is close to the one obtained in our work, $i\approx67^{\circ}$. Sales et. al. also determined the relation between the inner and the outer radii of the torus, $R_{0}/R_{d}=20$, where $R_{0}\sim11$\,pc. Additionally, in Fig.\,\ref{fig5} we show the 70-month Swift/BAT crab-weighted light curve for NGC\,3281 in 14--195\,keV energy range. As we can see the 14--195\,keV count rate is strongly variable on time-scales from months to years, showing a dynamic range of about a factor of $\sim5$ between the highest and the lowest count rate states. Thus our results, along with the results of the previous works confirm the existence of the Compton-thick clumpy dusty torus in this AGN.

\section*{\sc summary and conclusions}
\vspace*{-1ex}
\indent \indent We have analysed the broad-band 0.2--150\,keV X-ray properties of the active galactic nucleus of Sy2 galaxy NGC\,3281. The observational data from XMM-Newton and INTEGRAL satellites give us a possibility to study the absorption features, continuum components, as well as the iron and nickel fluorescent lines Fe\,K$_{\alpha}$ and Ni\,K$_{\alpha}$. The main results of this study can be summarized as follows:
\begin{itemize}
\item the obtained value of the hydrogen column density along the line of sight, $N_{H}=2.08\cdot10^{24}$\,cm$^{-2}$, confirms the Compton-thick nature of NGC\,3281 galaxy;
\item the continuum contains reflected component, possibly originated from the back-side reflection, and the zero-order continuum dominates over it in the spectrum above $\sim10$\,keV;
\item the fluorescent Fe\,K$_{\alpha}$ line demonstrates the stable flux level, as well as in the previous works \cite{RPB, Simpson, Tit, WMRT}.
\end{itemize}

In the same time the 70-month Swift/BAT lightcurve of this object demonstrates the irregular variability above 15\,keV, with the amplitude of $\sim5$. This can be considered as an evidence for a dusty clumpy torus. The spectral features in the mid-IR range also confirm the presence of a torus-like structure with Compton-thick clouds \cite{SPR, Simpson}.

We also found the inclination angle between the torus polar axis and the observer's line of sight $i\approx67^{\circ}$. This result is in a good agreement with the value obtained from the mid-IR observations \cite{SPR}.

Let us note that the continuum above $\sim70$\,keV can not be described only by toroidal reprocessor model plus comptonized spectrum. In the same time the $\sim 15-50$\,keV spectrum can not be described only by the disk-reflected model. Thus it would be very desirable to use the model containing the both.

The mild broadening of the Fe\,K$_{\alpha }$ ($\sigma\approx2-50$\,eV) and Ni\,K$_{\alpha}$ ($\sigma\approx106$\,eV) emission lines corresponds to their origin inside the gas/dust torus or in its inner part, or in the outer part of the broad line region. The values of EW for each line agrees with our $N_{H}$ value, according to the results of modelling of line production from \cite{MY,YM}.

Within the energy range 0.2--2\,keV the soft excess described by the thermal emission from the hot diffuse gas with $kT=590$\,eV was disclosed. Warm absorption is also present in the spectrum.

\section*{\sc acknowledgement}
\vspace*{-1ex}
\indent \indent We thank the anonymous referee for helpful and constructive comments and suggestions. This work has used data of observations obtained with XMM-Newton, an ESA  science mission with instruments and contributions directly funded by ESA Member States and NASA and data obtained from the High Energy Astrophysics Science Archive Research Center (HEASARC) provided by NASA's Goddard Space Flight Center.

\end{document}